\documentstyle[12pt]{article}
\begin{document}
\begin{titlepage}

\begin{flushright}
%\begin{minipage}
UAB-FT-338\\
May 1994
%\end{minipage}{12em}
\end{flushright}

\vspace{\fill}

\begin{center}
{\Large \bf Fixed boundary conditions and phase transitions
in pure gauge compact QED}\end{center}

\vspace{\fill}

\begin{center}
       {\large
        M. Baig and H. Fort\\
	Grup de F\'{\i}sica Te\`orica\\
	and\\
        Institut de F\'\i sica d'Altes Energies\\
	Universitat Aut\`onoma de Barcelona\\
	08193 Bellaterra (Barcelona) SPAIN\\}
\end{center}

\vspace{\fill}

\begin{abstract}
We have simulated the pure gauge compact QED with fixed boundary
conditions, on lattices from $6^4$ to $16^4$. We argue that a lattice
with this fixed boundary imposition corresponds actually to a lattice
with spherical topology.
We have found the presence of a phase transition without any
trace of discontinuity. Moreover, the specific heat and the Binder
cumulant are qualitatively consistent with a second order phase
transition. The implications of this observation on the nature of the
compact lattice QED are discussed.
\end{abstract}

\vspace{\fill}

\end{titlepage}
\newpage

\noindent{\em Introduction}
\bigskip

Numerical simulations of lattice gauge theories are actually
performed in a finite space-time box. The computer limitations imply
severe restrictions on the lattice size of this box. For example, the
largest lattices used today to simulate lattice  QED are of the order
of $22^4$.
To avoid the border ``effects'' generated by such a box, periodic
boundary conditions are usually adopted. From a geometrical point
of view, this fact implies that the system is actually simulated on an
hypertorus. In most cases this geometrical effect is not relevant, but
in some other cases the association of a finite-box with periodic
boundary conditions may originate unphysical effects.

Pure gauge lattice QED constitutes one of these special cases.
The initial measures\cite{ln} in small lattices suggested
the presence of a second order phase
transition but this transition became a first order one
when the simulations were performed on
larger lattices\cite{jnz}\cite{acg}.
In order to explain this fact one has to look at the confining
mechanism in lattice QED.

Some time ago Poliakov\cite{p} and Banks, Myerson and Kogut\cite{bmk}
showed that in the lattice (compact) QED monopole excitations appear
naturally.
Furthermore, the monopoles play a central role in the explanation of
the phase-structure.
In fact, they produce the ``disorder'' and give raise to the
confinement.
In $D=3$ monopoles are 0-dimensional pointlike excitations whilst
in $D=4$ they become 1-dimensional, and, due to the magnetic flux
conservation, they form closed loops. The density of monopole loops
in four dimensions was measured as a function of the
coupling $\beta$ and they observed a fall-off of the density over the
phase transition.

A possible explanation for the discontinuity in the internal
energy relies on the fact that
a lattice with periodic boundary conditions is a hypertorus, so
topologically nontrivial loops which wrap around the lattice
are permitted. This wrapped loops are present only in the
the confining side \cite{gup}\cite{sw} and clearly they have a fairly
large action associated with them. So a jump is observed when they
disappear at deconfining transition.
This jump no longer survives in the infinite
volume limit. In other words, the first order behaviour seems to be a
spurious topological effect occurring
in finite lattices with periodic boundary conditions
\cite{gup}\cite{gr}.

This idea is supported by the recent simulations of
the Wilson action on  ``closed topology'' lattices by Lang and Neuhaus
\cite{lan}. These simulations, being performed on the four-dimensional
boundary of a five-dimensional hypercube, found no metastability
signal at the phase transition point. They suggest that, contrarily to
the studies of compact QED on hypercubic lattices with periodic
boundary conditions, the phase transition of pure gauge QED becomes
of second order when simulated on lattices with the topology of a
sphere.
However, as pointed out recently in \cite{krw}, these simulations still
have some technical difficulties that can spoil the asymptotic scaling
regime.

This letter is intended to contribute to the clarification of the real
nature of the U(1) (compact) phase transition studying a
different implementation of the spherical topology.

\bigskip
\noindent{\em Fixed boundary conditions}
\bigskip

Our starting point is
the old observation\cite{ms} that in any local gauge theory the
Wilson loop average in an infinite lattice can be bounded from above
and below by the corresponding expectation value in a finite lattice
with appropriate boundary conditions. In fact, the lower bound
is reached with {\em free} boundary conditions and the upper bound
can be obtained with {\em fixed} boundary values (i.e. all the gauge
factors U(l) = 1 for l belonging to the boundary). This two bounds
coincide outside the transition region and if one imposes the standard
{\em  periodic boundary conditions} the average of the Wilson
loop interpolates between the two bounds exhibiting a kind of jump.

Looking carefully at the figures  of\cite{ms} (obtained actually from
rather small lattices), the behaviour of the fixed
boundary run of the pure gauge compact QED
shows a clear change of slope near the phase transition point. This
curve seems to represent more a true transition than to be
a simple {\em bound} of the internal energy. The main point is that
a simulation performed with a system with fixed boundary conditions
may corresponds actually to a system simulated on a lattice with
spherical topology.

A simple geometrical argument may clarify this point. Fig 1 shows a two
dimensional square lattice $L\times L$.
Imagine that all links over the boundary are put and fixed to unity
(dashed and dotted lines of Fig 1).
All plaquettes containing one of these links will have a contribution
to the action as a product of only three independent elements of U(1),
representing a loop as a triangle, instead of an elementary square.
This can be imagined, from a 3-d world point of view, as
collapsing to a single point all the square lines of the border, i.e.
obtaining a sphere instead of a torus.

\bigskip
\bigskip
\noindent{\em Numerical simulations}
\bigskip

To check this scenario we have performed a numerical simulation of
the compact pure gauge U(1) theory on a lattice with fixed boundary
conditions, studying the behaviour of the thermodynamical quantities
(internal energy, specific heat, and Binder cumulant). To understand
how we implemented these boundary conditions, let us to consider
again the two dimensional square lattice of Fig. 1.
We denote the points and links by
$$n(i,j),\ \ \  i,j=1,L,$$
and
$$l_\mu(i,j),\ \ \ \mu=1,2,$$
respectively.

To ``close'' the lattices one has to assign values for the ``extra''
links
$$l_2(L+1,j),\ \ \ j=1,L,$$
and
$$l_1(i,L+1),\ \ \ i=1,L,$$
i.e. the dotted links of Fig. 1.
If one imposes periodic boundary conditions, this assignment is
performed repeating the values of the links belonging to the border
$$l_2(1,j),\ \ \ j=1,L,$$
and
$$l_1(i,1),\ \ \ i=1,L,$$
i.e. the dashed links of Fig. 1.

Our implementation of the fixed boundary conditions takes into
account this fact.
We fix all dashed links to unity. Periodic boundary conditions
fix also all remaining border links (dotted) to unity obtaining,
actually, a lattice with spherical topology. Finally, one has only to
avoid the Monte Carlo updating of these links during the simulation.

Our simulation has been performed in four dimensions. The border of a
square domain in 4-d is a set of eight 3-d cubes.
All plaquettes contained in these cubes are put to unity.
Note that some plaquettes will have all their links fixed to unity.
Those plaquettes will remain unchanged along the simulation.
The effect of these plaquettes is merely to add
a constant factor to the lattice action and, hence, they have
no physical meaning.

The number of free plaquettes
(i.e. rejecting those duplicated by the periodic boundary
conditions) in a simulation is just
$$ N_p^t=6L^4,$$ but the number of free plaquettes (without
counting those that are fixed to unity) is just
$$N_p^f=6L^4-12L^3+6L^2.$$

This number comes from the fact that one has to subtract the plaquettes
belonging to the eight cubes of the boundary, taking into account that
some faces of these cubes are shared by them.
The knowledge of this factor is necessary to normalize the average
internal energy density taking into account only the non-fixed
plaquettes, but including all plaquettes with some link variables
(i.e. the "triangles" of the 2-d example).
This differs from the usual way of considering only
the innermost four dimensional cubes of the lattice\cite{ms}.

Note that with these fixed boundary conditions just one point of the
lattice has a special connectivity with his neighbours. The
consistence of our results seems to point out that the effect of
this point is not fundamental in the phase structure of the theory.

\bigskip
\noindent{\em Results}
\bigskip

Since we have only changed the implementation of the boundary
conditions, our lattice action is the usual compact U(1) action
\begin{equation}
S= \beta\sum_p(1-\cos\theta_p),
\end{equation}
where $\theta_p$ stands  for the circulation of the gauge field
around a plaquette and $\beta = 1/e^2$ is the gauge coupling.
We have measured the internal energy density
\begin{equation}
<E> = \frac{1}{N_p^f}\sum_p(1-\cos\theta_p),
\end{equation}
the specific heat, through the energy fluctuations,
\begin{equation}
C=N_p^f(<E^2>-<E>^2),
\end{equation}
and, finally, the Binder cumulant
\begin{equation}
B=1-\frac{<E^4>}{3<E^2>^2}.
\end{equation}

To perform the numerical simulations
we have chosen a standard heat bath simulation algorithm adapting an
existing and well tested  program\cite{cpc} to the new boundary
conditions.

A detailed finite size analysis of this model
is out of the scope of this letter and it will
require a considerable numerical effort.
Indeed, to fix the links of the 3-d border to unity  implies
that a significant fraction of the links is not changed and, hence,
we have to go to larger lattices.
The numerical results presented here come from some thermal cycles
performed for lattice sizes $6^4, 8^4, 10^4, 12^4, 14^4$ and $16^4$.
The statistics has been of 10.000 iterations per each $\beta$ value.

In Fig.2 we show the internal energy for different lattice sizes. No
sign of discontinuity is observed. Moreover, a cross-over point seems
to appear as L increases, just approaching $\beta =1$. This may be
compatible with the presence of a continuous phase transition.

In Fig.3 we plot the specific heat $C$. The emergence of a peak which
grows with $L$ and moves towards $\beta$ = 1 can be clearly observed.
Also the growing of that peak seems slow enough to ensure the absence
of a weak first order transition.

The Binder cumulant is plotted in Fig.4. It shows a clear
convergence to the expected value of 2/3 characteristic of a second
order transition. Furthermore, one can see that the value of $\beta$
for the minimum of the Binder
cumulant and that for the maximum of $c$, being different for small
lattice sizes, they coincide for $L \geq$ 14.

\bigskip
\noindent{\em Conclusions}
\bigskip

These results, although being basically qualitative, point out the
continuous character of the U(1) phase transition when periodic
boundary conditions and the spurious effects they bear with are
avoided.
The next step would be
a more detailed study (with much bigger statistics) in order to extract
the critical exponents and check if they verify the hyperscaling
relations. This will require a great computational effort since it
would be necessary to simulate much large lattices.

\bigskip
\noindent{\em Acknowledgments}
\bigskip

Numerical computations have been performed mainly on the cluster
of IBM RISC/6000 workstations of the IFAE.
 This work is partially supported by research
project CICYT number  AEN 93-0474 and CESCA (Centre de Supercomputaci\'o
de Catalunya).

\newpage

\noindent{\large Figure captions}

\begin{enumerate}

\item  Sketch of a two-dimensional square lattice.
Dotted links are the replica of
the dashed ones generated through periodic boundary conditions.
When imposing the fixed boundary conditions all these links are fixed
to unity giving an ``effective'' spherical topology.

\item  Internal energy measurement on lattice sizes from $6^4$ to $16^4$
with fixed boundary conditions. Results come from a thermal cycle with
10.000 measuring iterations per point.

\item  Specific heat measurement on lattice sizes from $6^4$ to $16^4$
with fixed boundary conditions.

\item  Binder cumulant measurement on lattice sizes from $6^4$ to $16^4$
with fixed boundary conditions. Dashed line represents the value 2/3.

\end{enumerate}

\end{document}